\documentclass[twocolumn,prb]{revtex4-2}
\usepackage[usenames,dvipsnames]{color}

\usepackage{comment}
\usepackage{graphicx}
\usepackage{inputenc}
\usepackage{lineno}
\usepackage{color}
\usepackage{hyperref}
\usepackage{amsmath}
\usepackage{braket}
\usepackage{caption}
\usepackage{bm}
\usepackage{subcaption}

\usepackage[]{todonotes}

\begin{document}
\title{Pairing susceptibility of the two-dimensional Hubbard model in the thermodynamic limit }
\author{Rayan Farid}
\affiliation{Department of Physics and Physical Oceanography, Memorial University of Newfoundland, St. John's, Newfoundland \& Labrador, Canada A1B 3X7} 
\author{Maxence Grandadam}
\affiliation{Department of Physics and Physical Oceanography, Memorial University of Newfoundland, St. John's, Newfoundland \& Labrador, Canada A1B 3X7} 
\author{J. P. F. LeBlanc}
\email{jleblanc@mun.ca}
\affiliation{Department of Physics and Physical Oceanography, Memorial University of Newfoundland, St. John's, Newfoundland \& Labrador, Canada A1B 3X7}

\date{\today}
\begin{abstract}
 We compute the diagrammatic expansion of the particle-particle susceptibility via algorithmic Matsubara integration and compute the correlated pairing susceptibility in the thermodynamic limit of the 2D Hubbard Model. We study the static susceptibility and its dependence on the pair momentum $\mathbf{q}$ for a range of temperature, interaction strength, and chemical potential. We show that $d_{x^2-y^2}$-wave pairing is expected in the model in the $U/t\to 0^+ $ limit from direct perturbation theory. From this, we identify key second and third-order diagrams that support pairing processes and note that the diagrams responsible are not a part of charge or spin susceptibility expansions.  We find two key components for pairing at momenta $(0,0)$ and $(\pi,\pi)$ that can be well fit as separate bosonic modes.  We extract amplitudes and correlation length scales where we find a predominantly local $(\pi,\pi)$ pairing and non-local $\mathbf{q}=(0,0)$ pairs and present the relative weights of these modes for variation in temperature, doping, and interaction strength.
\end{abstract}

\maketitle

\section{Introduction}
\label{main:intro}
Understanding the mechanisms by which electronic correlations drive phase transitions remains a dominant motivation for the development of numerical approaches for solving correlated electron systems.  The two-dimensional Hubbard model has been heavily studied in this respect\cite{benchmarks,Schaefer:2020} and is thought to be the quintessential representation of high transition temperature cuprate materials since the model, despite its simplicity, contains metallic, insulating, pseudogap and superconducting phases when studied by non-perturbative approaches on finite-sized systems.\cite{park:2008,gull:2012,gull:2013,chen:2015}  There the finite-size approach allows for a second-order phase transition to the superconducting state while a truly infinite system cannot due to the Mermin Wagner theorem.  Nevertheless, perturbative expansions at high order on infinite systems find corroborating physics in the model; metallic, insulating, and pseudogap features in single-particle properties without the need for infinite range correlations.\cite{fedor:2020,Schaefer:2020}  In all cases, the dominant excitations leading to insulating and pseudogap behaviour are antiferromagnetic $\mathbf{q}=(\pi,\pi)$ spin excitations and this has been shown explicitly via fluctuation diagnostics methods\cite{gunnarsson:2015,behnam:2020,wu:2017} and rigorously tested by a variety of numerical approaches.\cite{Schaefer:2020} To complicate matters, recent works\cite{gull:pnas:2022,gull:natphys:2022} have shown that the dominant excitation responsible for an anomalous self-energy is identical to the dominant excitation thought to lead to insulating behaviour, primarily static $(\pi,\pi)$ spin-excitations.

For the general problem, particle pairing is governed by a Bethe-Salpeter equation for the particle-particle susceptibility $\chi_{pp}=\chi_0+\chi_0 \Gamma \chi_0$ where $\chi_0$ is given by the dressed particle-particle bubble and all pair-correlations are described by the insertion of the full vertex $\Gamma$.  Correlated pairing processes are therefore a property of only vertex diagrams, and in the case of the Hubbard model, these have been somewhat studied via non-perturbative approaches by computing the correlated pairing susceptibility $P_g(\mathbf{q},\Omega)$ that is nothing more than the vertex contributions projected into the $g=s,p,d...$ channels.\cite{Maier2019,sigrist:1991,Scalapino12,tsuei:2000} 
Existing studies of the pairing susceptibility are from non-perturbative approaches such as the dynamical cluster approximation that have very limited momentum space resolution.\cite{chen:2015,scalapino:2007}   From perturbative expansions very little is known since the simplest vertex terms, the RPA-like ladder series does not give rise to divergent behaviour when projected into the $d$-wave channel. The application of a form factor strongly suppresses the ladder diagrams when scattering momenta $\mathbf{q}$ is along its nodal line rendering the ladder expansion ineffectual.

In this work, we directly compute the full perturbative expansion of the vertex contributions to the particle-particle susceptibility.  This is possible due to algorithmic advances that provide partially analytic expressions to arbitrarily complex Feynman diagrams without finite size approximations.\cite{AMI,libami,AMI:spin,GIT,mcniven:2022,mcniven:2021} We study the full $\mathbf{q}$-dependence of the static pairing-susceptibility and impose a variety of symmetry factors to extract the susceptibility of the system to $s$-wave, $p$-wave, and $d$-wave symmetries.  Our results are extraordinarily consistent with previous works, both  and non-perturbative, and we identify the primary components of pairing.

\section{Models and Methods}
\subsection{Hubbard Hamiltonian}
\label{main:Hubbard}
We study the single-band Hubbard Hamiltonian on a 2D square lattice\cite{benchmarks},
\begin{eqnarray}\label{E:Hubbard}
H = \sum_{ ij \sigma} t_{ij}c_{i\sigma}^\dagger c_{j\sigma} + U\sum_{i} n_{i\uparrow} n_{i\downarrow},
\end{eqnarray}
where $t_{ij}$ is the hopping amplitude, $c_{i\sigma}^{(\dagger)}$ ($c_{i\sigma}$) is the creation (annihilation) operator at site $i$, $\sigma \in \{\uparrow,\downarrow\}$ is the spin, $U$ is the onsite Hubbard interaction, $n_{i\sigma} = c_{i\sigma}^{\dagger}c_{i\sigma}$ is the number operator.  We restrict the sum over sites to nearest and next-nearest neighbors for a 2D square lattice, resulting in the free particle energy 
\begin{eqnarray}
\nonumber\epsilon(\textbf k)=-2t[\cos(k_x)+\cos(k_y)]-4t^\prime [\cos(k_x)\cos(k_y)]-\mu,
\end{eqnarray} 
where $\mu$ is the chemical potential, and $t$($t^\prime$) is the nearest (next-nearest) neighbor hopping amplitude. Throughout we work with energies in units of the hopping, $t=1$, and primarily study the $t^\prime=0$ case.

\subsection{Pair Correlation Function}
\label{main:Pg}
Although a second-order phase transition to a superconducting state is not allowed in two dimensions it remains possible that fluctuations to the anomalous Green's function can be created in the presence of an external pair-generating field. We define the anomalous Green's function $F(\mathbf{q,\tau})= -\langle T c_{\mathbf{k+q}\uparrow}(\tau) c_{-\mathbf{k}\downarrow}(0)\rangle$ which relates to a specific particle-particle susceptibility  theory via linear response theory 
\begin{equation}\label{E:linear}
 \int_{0}^{\beta}\chi_{pp\overline{\uparrow\downarrow}}(q,\tau) e^{i\Omega \tau} d\tau  = \int_{0}^{\beta} \frac{dF(q,0)}{d\eta(q,\tau)} \bigg\rvert_{\eta \to 0}  e^{i\Omega \tau} d\tau
\end{equation}
 where $q$ is the scattering momenta, $T$ is the time-ordering operator and $\eta(q,\tau)$ is strength of the generating field.\cite{chen:2015} Interpretation of Eq.~(\ref{E:linear}) is that a positive value of $\chi_{pp\overline{\uparrow\downarrow}}$ will relate to a positive change in the superconducting order parameter for application of a superconducting field in the linear response regime.  Should this susceptibility diverge then a superconducting order with infinite range will have formed.
  One can project  Eq.~\ref{E:linear} to the irreducible representation of the square lattice  and  derive the projected susceptibility as
 \begin{equation}\label{E:Pg}
    \chi^{g}_{pp\overline{\uparrow\downarrow}}(\mathbf{q},\tau)=\langle \gamma_{k+q} c_{\mathbf{k}+\mathbf{q}\uparrow}(\tau) c_{-\mathbf{k}\downarrow}(\tau) c^{\dagger}_{-\mathbf{k'}\downarrow}(0)c^{\dagger}_{\mathbf{k'}+\mathbf{q}\uparrow}(0)\gamma_{k'+q}  \rangle. 
\end{equation}
 This observable identifies the transition to a superconducting state and  quantifies the pairing process between a spin-up electron with momentum $\mathbf{k}+\mathbf{q}$ and a spin-down with momentum $\mathbf{-k}$.  In the BCS limit, the susceptibility is expected to be dominated by a zero momentum pair given for $|q|=0$. 
  
The subscript $g$ represents the choice of $s,p,d-$wave projection of $D_{4h}$  symmetry group  and determines the functions $\gamma_{\mathbf{k}+\mathbf{q}}$ and $\gamma_{\mathbf{k^\prime}+\mathbf{q}}$ that represent initial and final momenta along a single particle line. We restrict ourselves to single component order parameters in the first harmonics of each representation (see appendix \ref{appendix:symmetry}).
 A second-order phase transition to a superconducting state is attained  when $\chi^{pp}_{\overline{\uparrow\downarrow}}$ diverges. The Bethe-Salepeter equation for the particle-particle susceptibility takes the form 
\begin{equation}
    \chi_{pp\overline{\uparrow\downarrow}} = \frac{\chi_{0} }{1+ \Gamma_{pp\overline{\uparrow\downarrow}}\chi_{0}}
\end{equation}
and  when the eigenvalue of the vertex component $-\Gamma_{pp\overline{\uparrow\downarrow}}\chi_{0}$ approaches unity, superconductivity for the symmetry of the eigenfunction is realized.

 Since the transition is only an attribute of the vertex correction $\Gamma$, divergence to the pair correlation  function $P_{g} = (\chi_{pp\overline{\uparrow\downarrow}} - \chi_{0})_{g} $ can be used as an indicator for the proximity to superconductivity.  While the eigenvalue may cross unity for a finite-sized system and exhibit a long-range order for a non-zero $T_{c}$ due to divergent susceptibility, in the 2D thermodynamic such behavior is \emph{never} allowed unless at  $T=0$. Nevertheless, one may still study the evolution of  $P_{g}$ over a range of parameter space, determine the tendency for single component order formation, and extract the relevant length scales. 

It is well established that the 2D Hubbard model has dominant charge and spin excitations.  For the infinite system, those excitations in the particle-hole channel remain finite length for non-zero temperatures.  Hence we expect finite range antiferromagnetic fluctuations and incommensurate charge excitations to coexist with the pair fluctuations we identify in this work.  Charge and spin susceptibilities have been extensively studied \cite{mcniven:2022,gull:2020:charge,Schaefer:2020,maier:2022,maier:2022:2} and do not conflict, compete, nor relate to the pair correlations we present.

\subsection{Diagrammatic Expansion}
\label{main:diagrammatic}
\begin{figure}
    \centering
    \includegraphics[width=\linewidth]{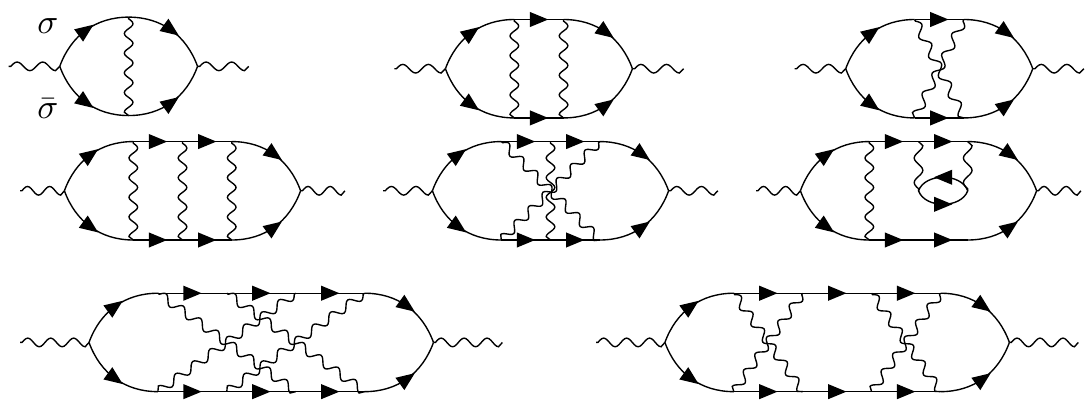}
    \caption{\label{fig:Diagra} A subset of Feynman diagrams responsible for the  vertex correction  to pairing. Straight lines with  arrowhead and wavy lines correspond  to single particle fermionic propagators and  onsite interactions $U/t$ respectively. The number of wavy lines determines the order. Onsite interaction must be between opposite spin  as a consequence of Pauli's exclusion principle $\sigma\neq\overline{\sigma}\in \uparrow\downarrow $}. 
\end{figure}

We perform a fourier transform to $P_{g}$ in Eq.~\ref{E:Pg} which  becomes a function of the momentum difference $\mathbf{q}$ and  external bosonic matsubara frequency $\Omega$.  We evaluate this expectation value via a perturbative expansion for the Hubbard interaction, and this gives rise to a set of bare Feynman diagrams, a subset of which is depicted in Fig.~\ref{fig:Diagra}. These are the set of all vertex diagrams that contribute to the correlated pairing susceptibility $P_g(\mathbf{q},\Omega)$.\cite{Maier:2011,Scalapino12, Rohringer12} Throughout we will study the static quantity, setting the external bosonic line to zero $P_g(\mathbf{q})=P_g(\mathbf{q},\Omega=0)$. 

Making no assumptions about the topology of the diagram, each Feynman diagram is converted to  analytic expressions written as 
\begin{align}
&\frac{1}{\beta^n}\prod\limits_{i}^{n_v}V(q_i)\sum\limits_{\{k_n\}} \sum \limits _{\{\nu_n\}}\prod
\limits_{j=1}^N G^j(\epsilon ^j, X^j ) = \prod\limits_{i}^{n_v}V(q_i)\sum\limits_{\{k_n\}} I^{(n)}, \label{eqn:full}\\
& I^{(n)}=\frac{1}{\beta^n}\sum \limits _{\{\nu_n\}}\prod
\limits_{j=1}^N G^j(\epsilon ^j, X^j ), \label{eqn:goal}
\end{align}
where $n_v$ is the order or the number of interaction lines with amplitude $V(q_i)=U$ for the Hubbard interaction.\cite{Rohringer12}  $n$ 
is the number of summations over Matsubara frequencies $\{\nu_n\}$  and 
internal momenta $\{k_n\}$, and $N$
is the number of internal lines representing bare Green's functions $G(\,X)=1/(X-\epsilon)$ where $X$ is a linear combination of frequencies and $\epsilon$ is a linear combination of energies.

The functions to be integrated are too complex to write by hand as the Matsubara sums generate a large number of analytic terms. Therefore we utilize Algorithmic Matsubara integration (AMI)\cite{AMI,AMI:spin,libami} that resolves the Matsubara sums in Eq.~\ref{E:Pg} to generate $I^{(n)}$ symbolically via repeated application  of residue theorem.  This represents fully one-third of the internal integrations being exact to machine precision with virtually zero computational expense.  The remaining spatial integrals are performed using integration methods for continuous functions. Hence the computation is dominated by the expense of a sequence of nested integrals over internal momenta $\{k_n\}$ on the right-hand side of Eq.~(\ref{E:Pg}).  We enumerate the number of such diagrams and the number of analytic expressions generated via AMI at each order in Table \Ref{tab:nterms}. 
\begin{table}[h]
    \centering
    \begin{tabular}{c|c|c}
        Order, $m$ & Diagrams & Terms\\ \hline \hline
        0 & 0  & 0 \\
         1& 1& 4\\
         2 & 2 & 28 \\
         3 & 13 & 702\\
         4 & 74 & 16666  \\
         5 & 544& 559812 \\
    \end{tabular}
    \caption{\label{tab:nterms}Number of diagrams for the Hubbard interaction at each order $m$ and number of terms generated through Matsubara sums via the AMI procedure the particle-particle vertex expansion $P_{g}$.}   
\end{table} 
\subsection{Establishing the range of $U/t$ validity}
\label{main:U_valid}
While AMI can produce analytical expressions of internal Matsubara sums for an arbitrarily large order with minimal computational expense, factorial growth in the diagrams and momentum integral space necessitates truncation at some finite order and study weak coupling limits such that higher order corrections are small. The advantage of this approach is that there is no finite-size approximation and the results are already in the thermodynamic limit. Consequently, there will always exist a region of the phase diagram (small $U/t$, high temperature, heavily doped) where the perturbative expansion remains controlled and results are virtually exact. However, the perturbative expansion can become uncontrolled in the vicinity of half-filling where the contribution is usually large, and it is thus necessary to find a range of $U/t$ where $P_{g}$ remains a good estimate.

\begin{figure}
    \centering
    \includegraphics[width=\linewidth]{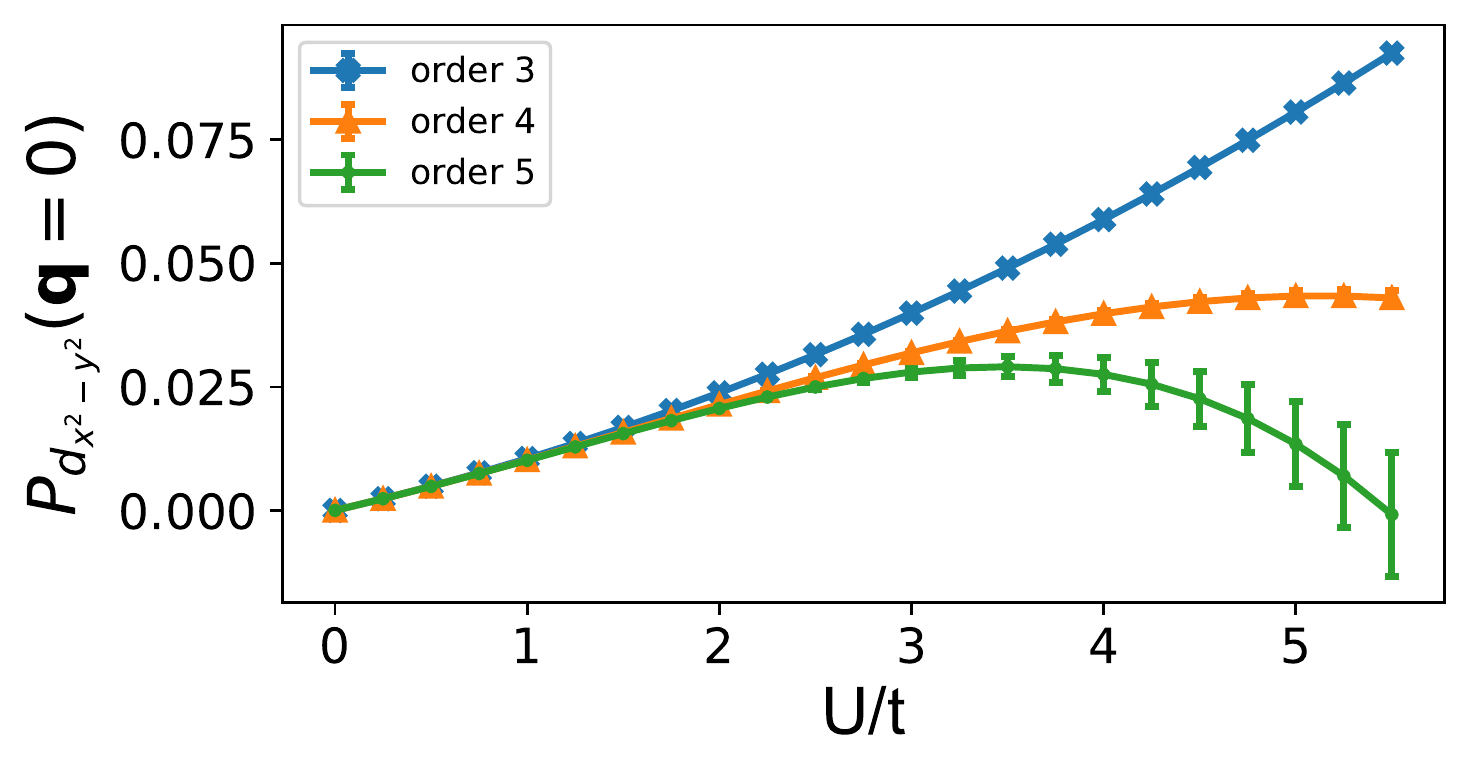}
    \caption{\label{fig:Ures1} Zero momenta  $d_{x^2-y^2}$-wave pairing truncated at  third, fourth, and fifth order as a function of interaction strength  $U/t$ at $\beta t =5 $ and half filling }.
\end{figure} 

To establish the region of validity, we proceed in an order-by-order expansion in powers of $U/t$ starting from small $U/t$ values for $d_{x^2-y^2}$-wave symmetry. We draw comparisons among truncated third, fourth, and fifth order $P_{d_{x^2-y^2}}$ at half-filling and find that within the $0<U/t<3$ range third order expansion  remains a good qualitative estimate. Beyond $U/t = 3$ fourth and fifth order corrections have an increasingly dominant effect on truncated third order as illustrated for one $\beta t =5$ case in Fig.~\ref{fig:Ures1}. Higher-order contributions are larger at
lower temperatures, indicating that our truncated third-order perturbative expansion remains valid only within the weak
coupling $U/t<3$ limit and intermediate temperature ranges.


\section{Results}
\subsection{ Order by Order Comparison}
\label{main:order}
\begin{figure}
    \centering
    \includegraphics[width=1.0\linewidth]{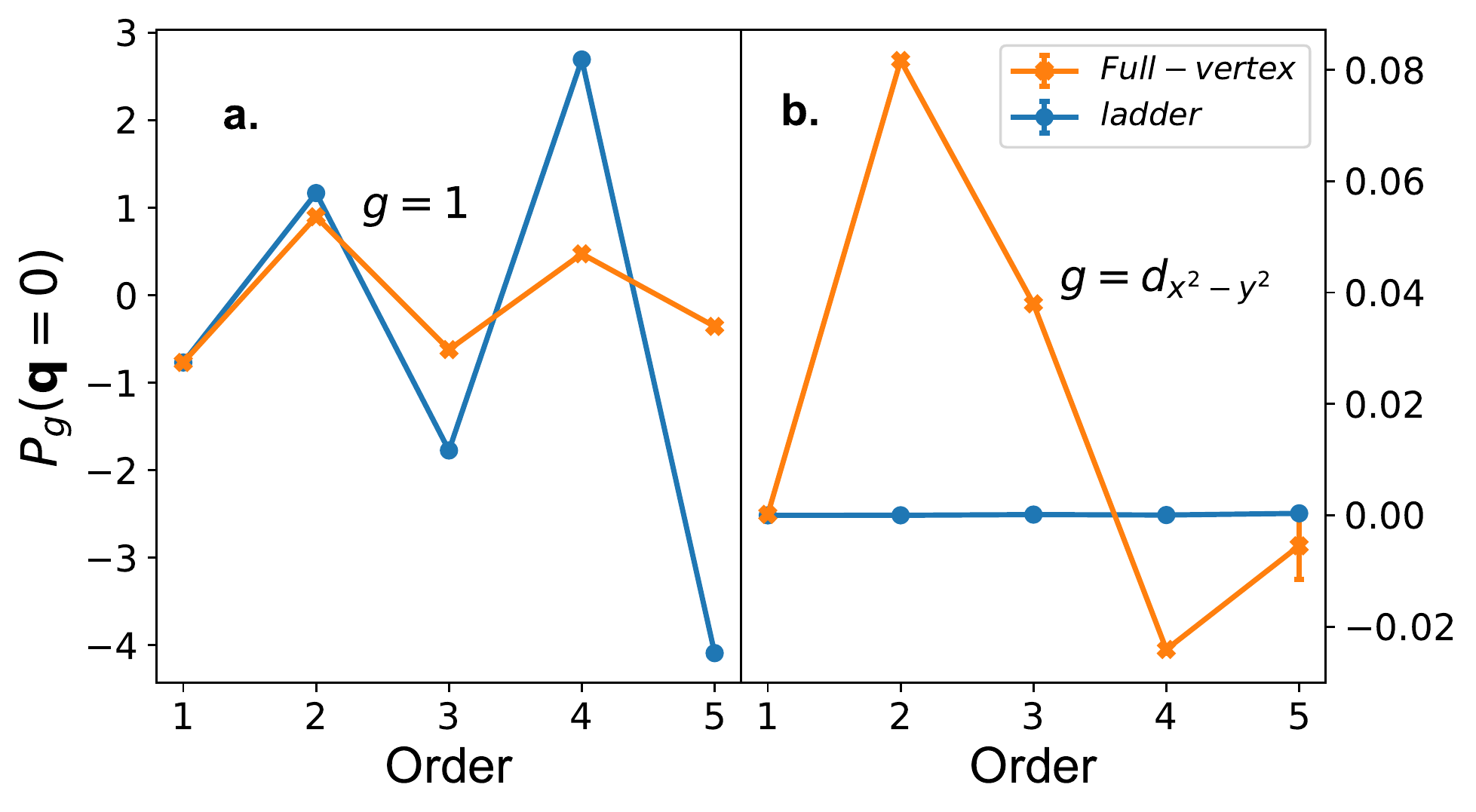}
    \caption{\label{fig:ordbyord}Order-by-order contribution to static pairing susceptibility for full vertex and ladder components  at half filling, $\beta t=5$, and $U/t=3.0$ for two symmetry factors: (a)  $S$-wave and (b) $d_{x^2-y^2}$-wave}
\end{figure}

We present the order-by-order breakdown of pair correlation function $P_{g}$ up to fifth order in Fig.~\ref{fig:ordbyord} for the half-filled case at $\mathbf{q}=(0,0)$, $\beta t=5$, and $U/t=3$ for $s-$ and $d_{x^2-y^2}-$wave symmetries.  We stress that in 5th order the results represent the sum of 544 diagrams that together are comprised of 559812 analytic terms to be integrated.  We contrast the contributions from all diagrams to contributions from only ladder diagrams - those accessible analytically by hand.  In the case of $s-$wave, Fig.~\ref{fig:ordbyord}(a), the dominant first-order ladder diagram is negative and opposes pairing.  The contributions then alternate sign order-by-order, but the negative value of the first order diagram makes s-wave pairing impossible in the $U/t\to0^+$ limit. Conversely, in the attractive interaction $U<0$ regime, the pairing is predominately an $s$-wave $\mathbf{q}=(0,0)$ process driven by ladder diagrams.  We see also that the non-ladder diagrams act to suppress the ladder diagrams at each order but do not change the qualitative behavior.  The $d_{x^2-y^2}-$wave case in Fig.~\ref{fig:ordbyord}(b) is entirely different.  Here the $\gamma_{\mathbf{k}}$ projection factors become independent of each other for ladder diagrams in the $|\mathbf{q}|\to0$ limit and produce zero contribution to pairing at every order.  Hence, the ladder diagrams neither promote nor \emph{oppose} $d_{x^2-y^2}-$wave pair formation, and this results in a dominant and positive second-order pairing susceptibility. This provides the potential for pairing even in the $U/t\to0$ limit.  Here shown for the half-filled problem, we find also positive contributions for the third order followed by competition at fourth order and a small negative 5th-order contribution. The sequence of positive second and third-order terms gives rise to a substantial range of $U/t$ where positive contributions exist before being squashed by higher-order terms.  We provide additional details and calculations on the order-by-order breakdown and pairing in $p$ and $d_{xy}-$wave channel in appendix \ref{appendix:kcut}-\ref{appendix:mu}.

\subsection{Momentum and Doping Dependence}\label{main:kandmu} 
\begin{figure}
    \centering
    \includegraphics[width=1.0\linewidth]{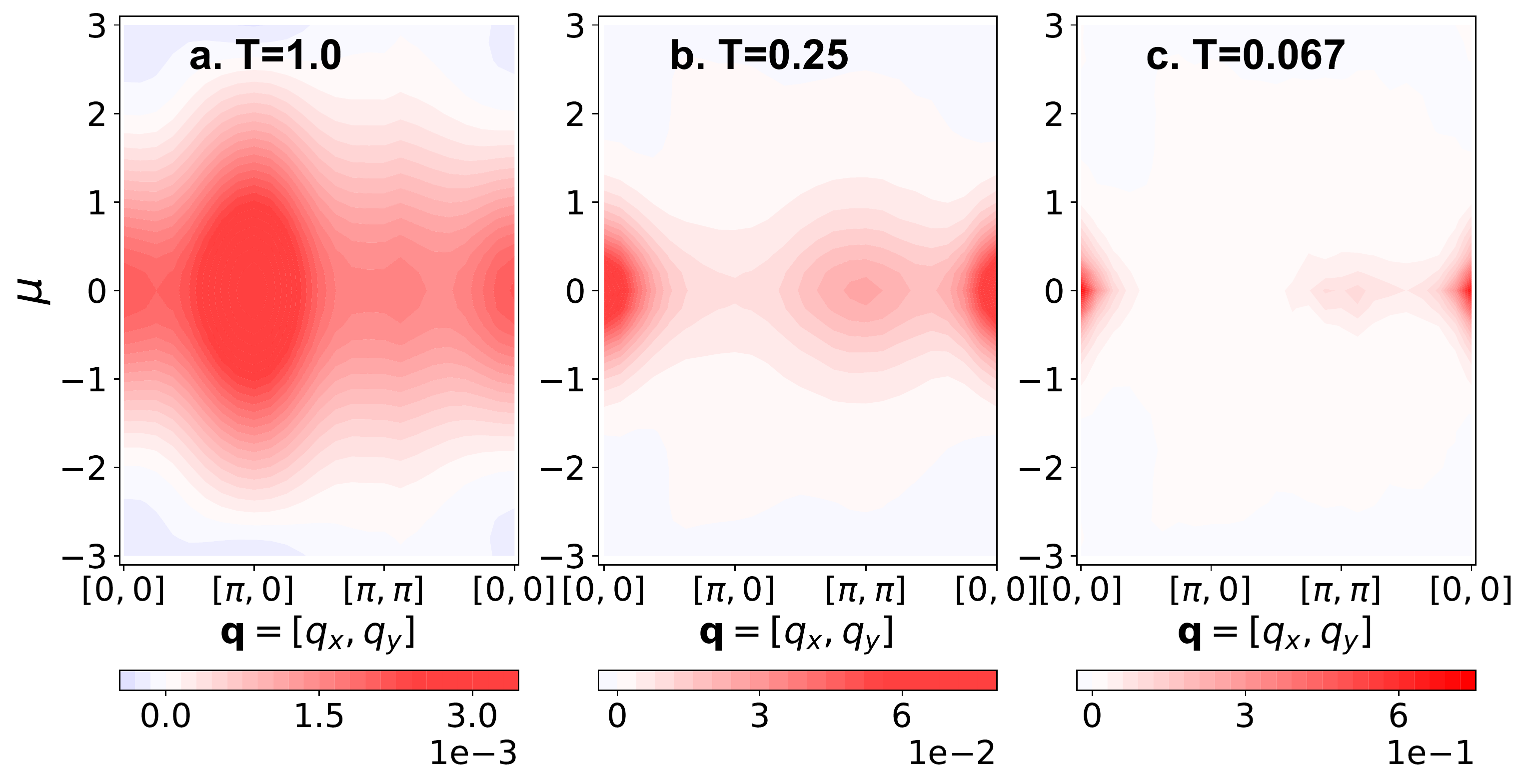}\\%
    \vspace{-0.085cm}\includegraphics[width=1.0\linewidth]{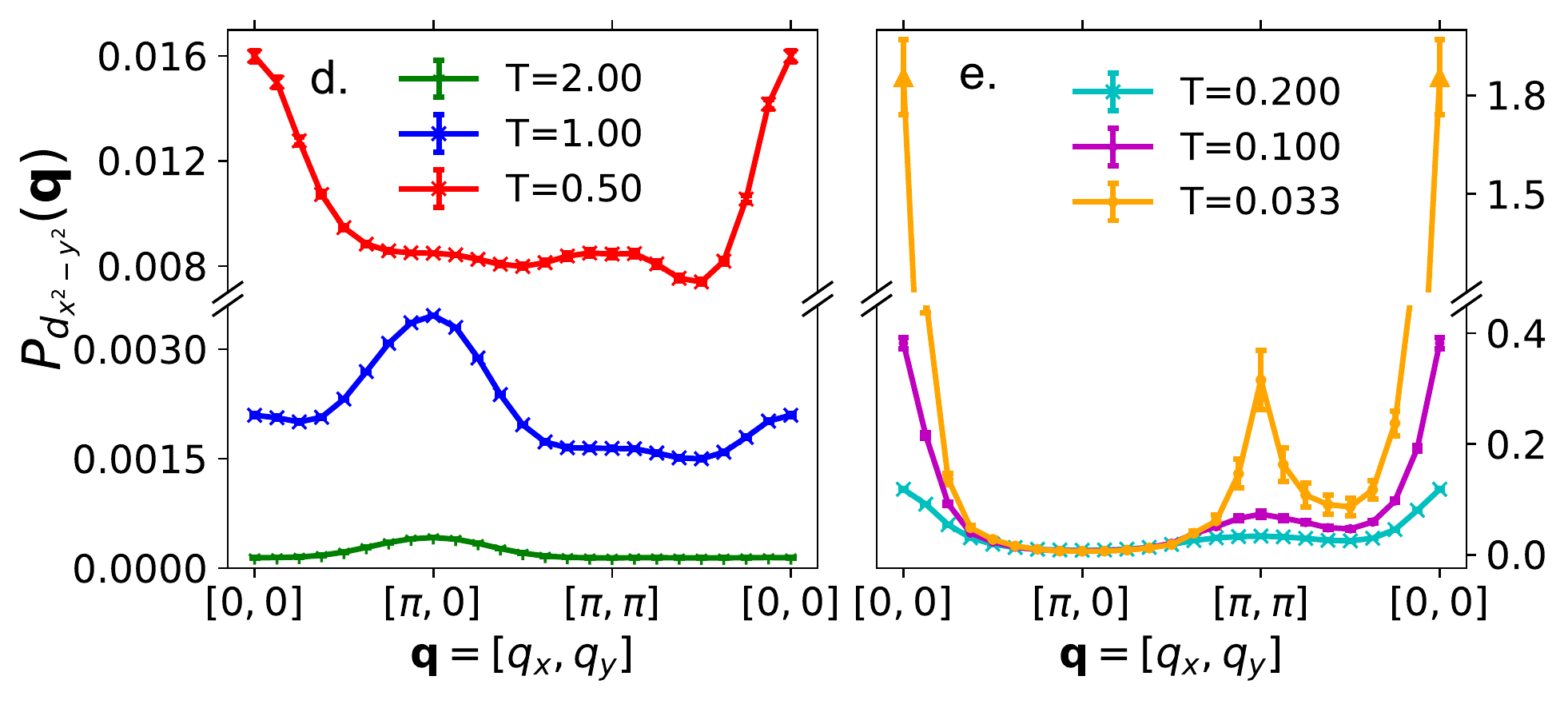}\\%
    \caption{ \label{fig:T_evol} (a-c) Contour plot for $d_{x^2-y^2}$ pairing across momentum symmetry cuts as a function of doping at $U/t=3$ for (a) $T=1.0t$, (b) $T=0.25t$, (c) $T=0.067t$. (d-e) the momentum evolution of pairing for various temperatures at $\mu=0$ and $U/t=3$    }
\end{figure}

In order to focus on the dominant positive contributions to the $d_{x^2-y^2}$ pairing that appears at second and third order we truncate the expansion at third order which allows us to quickly sweep through density and the full momentum dependence which we depict in the top row of Fig.~\ref{fig:T_evol}.  The three panels are for high, intermediate, and low temperatures at $U/t=3.0$ as a function of high symmetry cuts in scattering momentum $\mathbf{q}$ and variation in chemical potential.  At high temperatures, the susceptibility amplitude is weak, is centered near the half-filling point at $\mu=0$, and appears most robust near the momentum vector $\mathbf{q}=(\pi,0)$.  As the temperature decreases we see that by $\beta t=4$ much stronger features emerge at $\mathbf{q}=(0,0)$ and surprisingly at $\mathbf{q}=(\pi,\pi)$.  Decreasing temperature further the $\mathbf{q}=(\pi,\pi)$ peak is suppressed and the pairing is dominated by zero momentum pairing as expected.  This evolution in pairing occurs over a range of temperatures where a variety of processes are known to take place in the 2D Hubbard model such as the presence of a pseudogap and also a metal-insulator crossover.\cite{fedor:2020,Schaefer:2020}

We emphasize the temperature dependence of the color-plot by presenting the $\mu=0$ cuts in the lower row of Fig.~\ref{fig:T_evol}.  At high temperatures, the amplitude is extremely weak (on the scale of $10^{-3}$) with a peak near $\mathbf{q}=(\pi,0)$.
As temperature decreases the dominant features are that of the second-order diagrams (see appendix D) that already show two separate peaks at $\mathbf{q}=(0,0)$ and $(\pi,\pi)$. Decreasing temperature further we see these peaks become sharper while the peak height at zero momentum is dominant.

\subsection{Role of Second Nearest Hopping} \label{main:tprime} 
\begin{figure}
    \centering
    \includegraphics[width=1\linewidth]{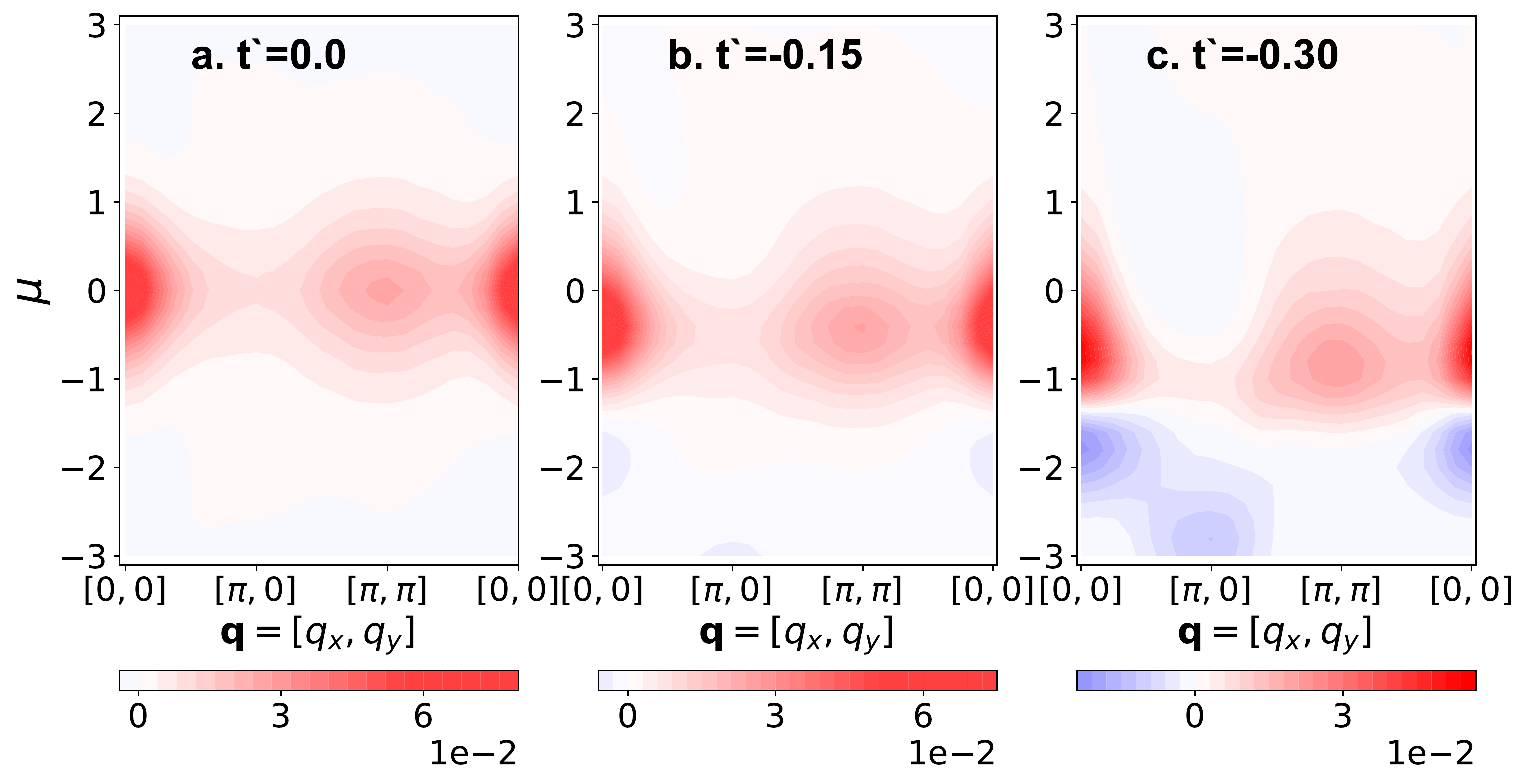}\\%
    \includegraphics[width=1\linewidth]{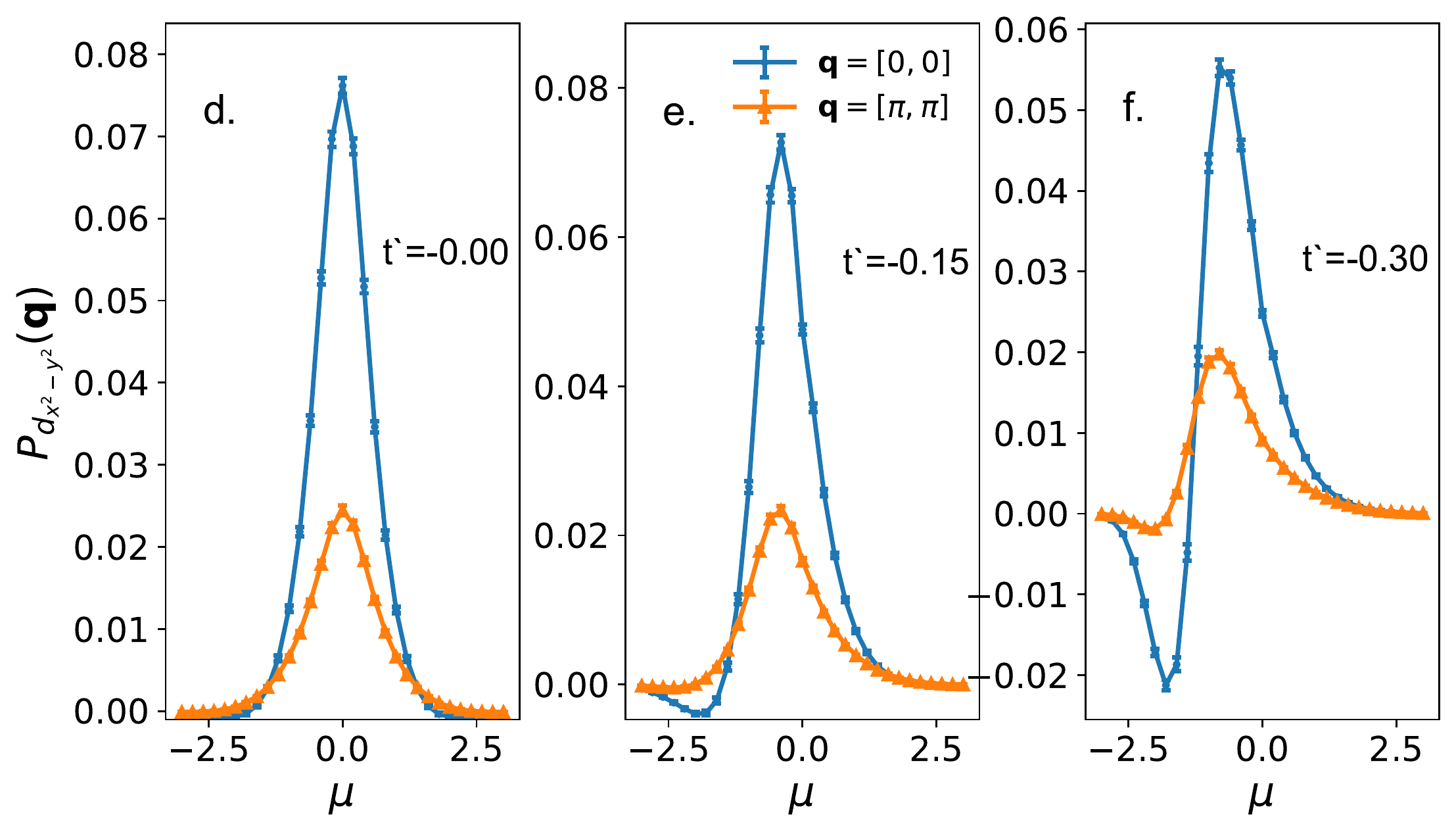}
    \caption{\label{fig:mu_q_map} (a-c)  Color plot for $d_{x^2-y^2}$ pairing across momentum symmetry cuts as a function of doping at $U/t=3$ and $\beta /t =4$ for $t^\prime=0.0$, $-0.15$, and $-0.3$.  (d-f) Competition between $\mathbf{q}=(0,0)$ and $\mathbf{q}=(\pi,\pi)$ pairing as a function of chemical potential for similar $t^\prime$ values. }
\end{figure}

We limit ourselves to two modest next-nearest neighbor hopping values $t^\prime = -0.15, -0.30$ on the scale of $t^\prime$ values used to fit the Fermi surface of cuprate materials from ARPES experiments. The effect of $t^\prime$ is that it shifts the single particle van Hove singularity to $\mu=4t^\prime$. Shown in Fig.~\ref{fig:mu_q_map}, the addition of $t^\prime$ appears to shift all peak features towards negative $\mu$ values, but does not directly follow the van Hove singularity. When focusing on the peak feature with increasing $t^\prime$, we notice a slight attenuation of pairing susceptibility. However at fixed $\mu$ values the inclusion of $t^\prime$ results in a dramatic reduction relative to the $t^\prime=0$ case. 
There exists also a range in $\mu$ below which the $\mathbf{q}=(0,0)$ susceptibility is negative. 

 We contrast the effect of $t^\prime$ for $\mathbf{q}=(0,0)$ and $\mathbf{q}=(\pi,\pi)$ and find that the $\mathbf{q}=(\pi,\pi)$ mode is somewhat less susceptible to the effect of $t^\prime$ as a function of doping.  It is worth noting that when $t'=-0.3$, Fig.~\ref{fig:T_evol}(f), it is possible to have a region where only  $\mathbf{q}=(\pi,\pi)$  susceptibility is attractive while the  $\mathbf{q}=(0,0)$ susceptibility becomes repulsive. 

\subsection{Fitting a two-component pairing}\label{main:Fit}  
\begin{figure}
    \centering
   \hspace{-0.3cm}\includegraphics[width=0.9\linewidth]{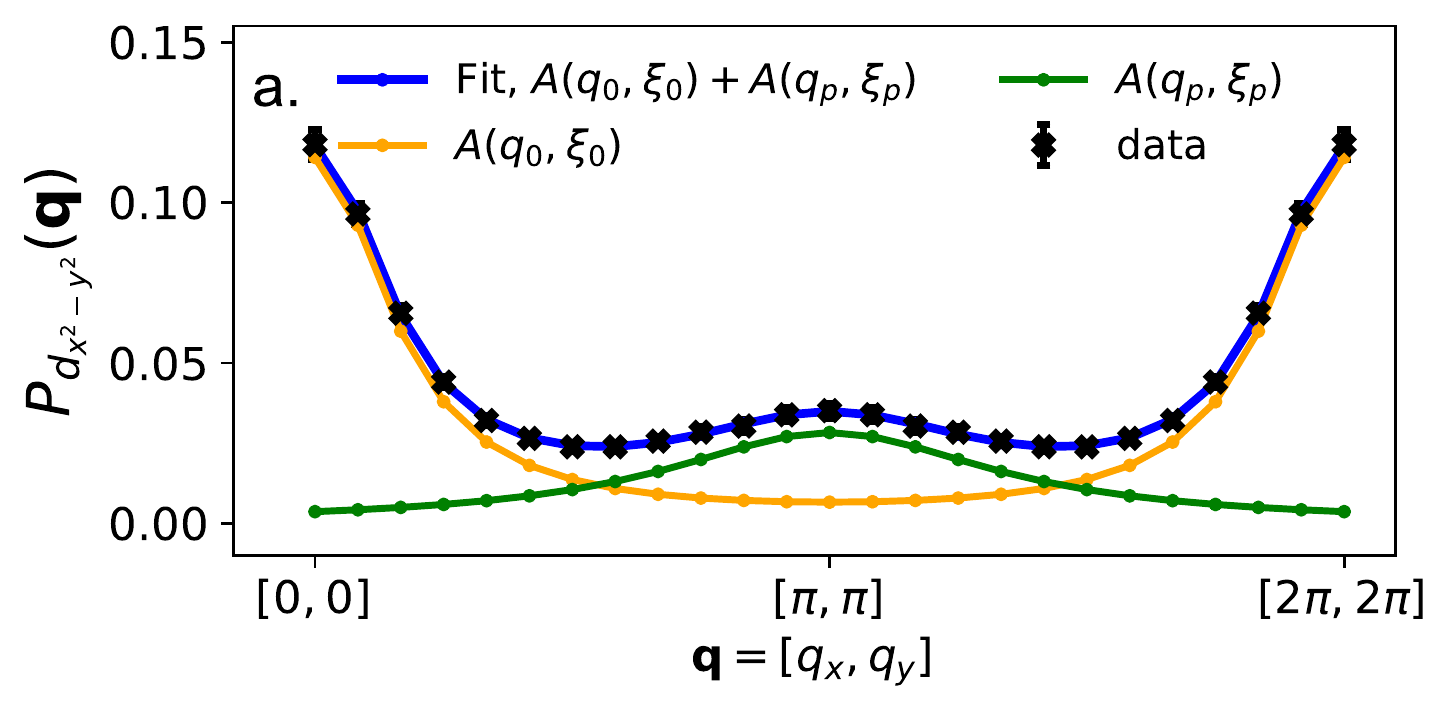}\\%
    \vspace{-0.1cm}\includegraphics[width=1\linewidth]{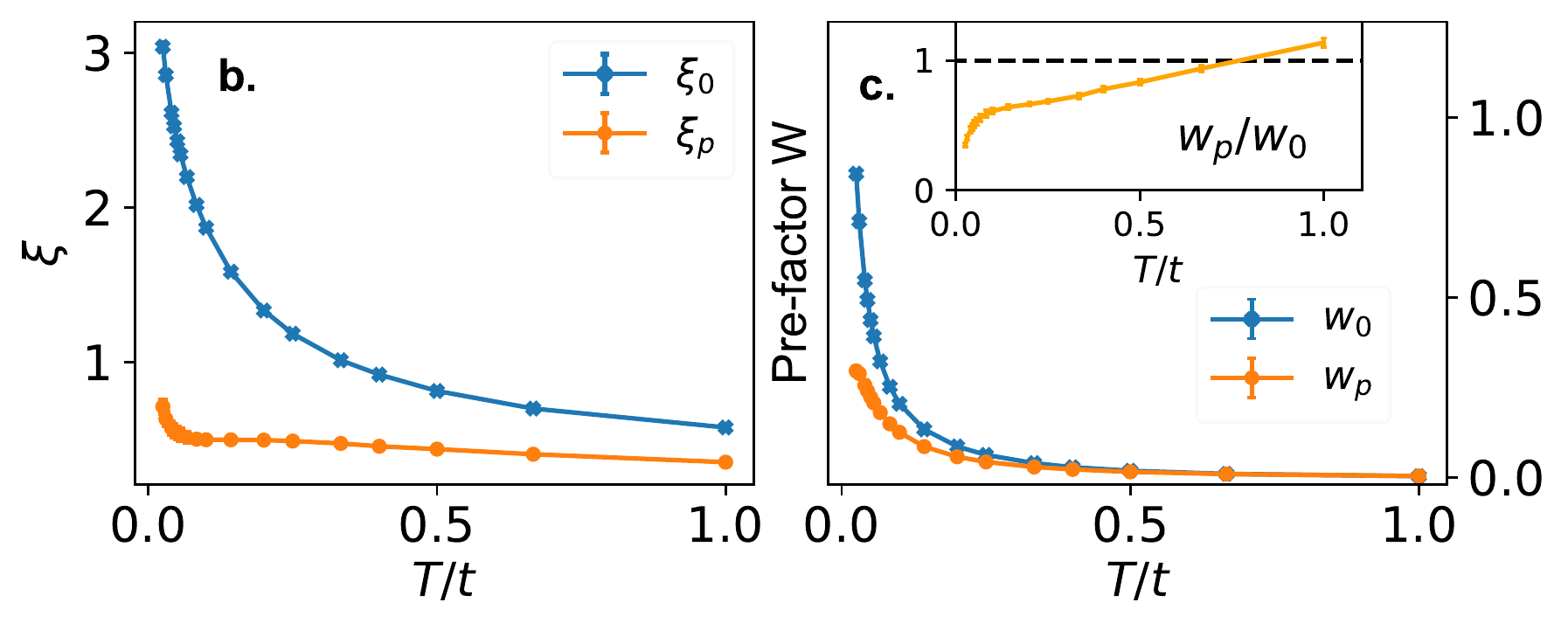}\\%
    \vspace{-0.17cm}\includegraphics[width=1\linewidth]{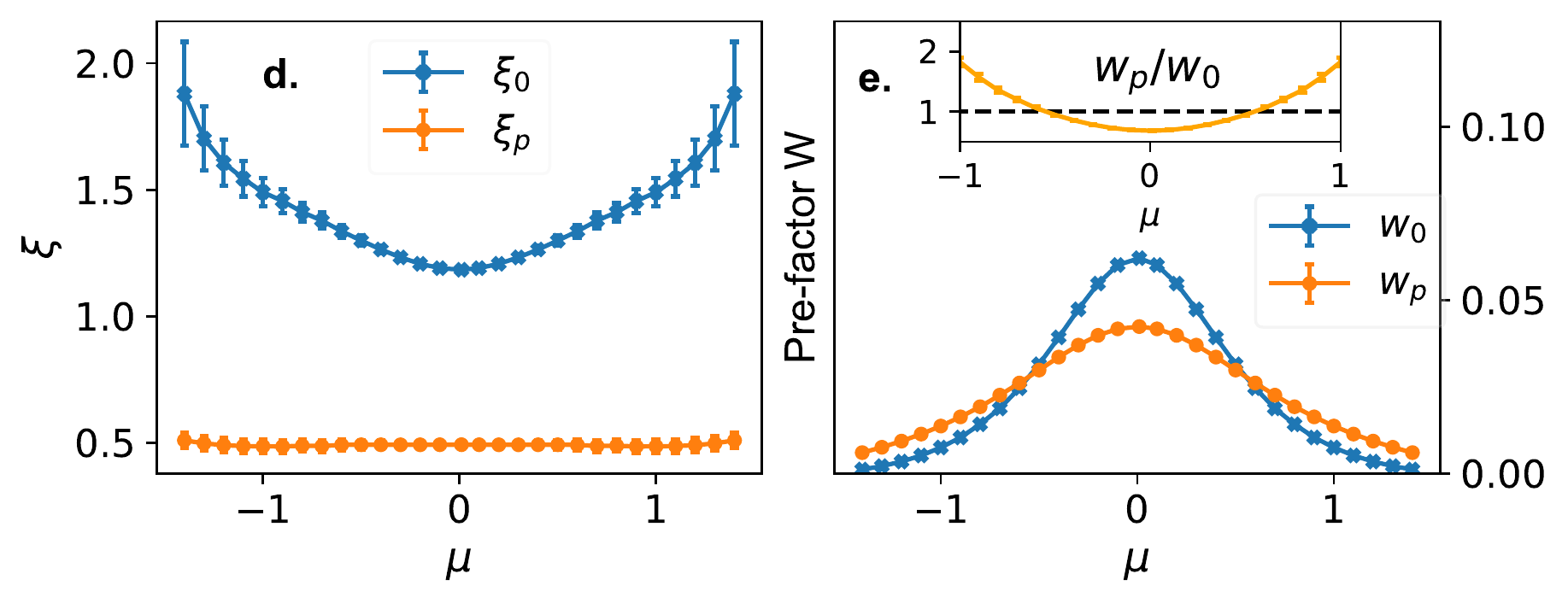}\\%
     \vspace{-0.18cm}\includegraphics[width=1\linewidth]{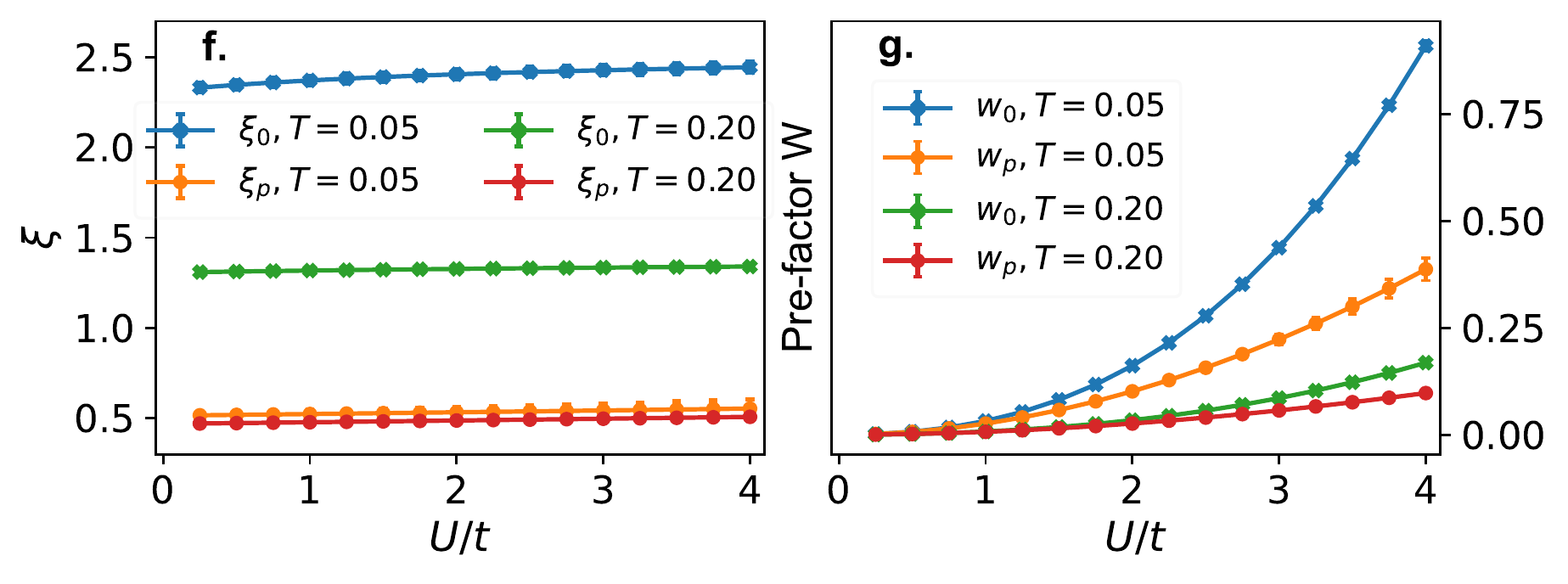}
    \caption{ \label{fig:xi} (a) Example fitting via Eq.~(\ref{eq:fit}).  (b-g) extracted correlation lengths ($\xi_0$ and $\xi_p$) in the units of lattice constant and weights ($W_0$ and $W_p$) 
    (b,c) the temperature dependence at $\mu=0$ and  $U/t=3$. (d,e) doping dependence at $T=0.25$ and  $U/t=3$. (e,f) interaction strength $U/t$ dependency at $\mu=0$ for temperatures $T=0.20t$ and $T=0.05t$.  Figure insets show the ratio between $W_{p}$ and $W_{0}$} .
\end{figure}

There exists a range of temperature and $U/t$ values where the result of $P_{d_{x^2-y^2}}$ can be precisely fit by two Lorentzian functions, one centered at $\mathbf{q}=(0,0)$ and a second at $\mathbf{q}=(\pi,\pi)$.  We emphasize that this is not apparent in the analytics of the diagrammatic expansions that when truncated at third order include the sum of 730 analytic terms each integrated over a $2(n+1)$ dimensional space at order $n$.  We define a fitting function  as a sum of two Lorentzian $A(\mathbf{q}_{0},\xi_{0}) +  A(\mathbf{q}_{p},\xi_{p})$ where
\begin{equation}\label{eq:fit}
\begin{split}
A(\mathbf{q}_{0},\xi_{0}) &= \frac{W_{0}\xi_{0}^{-1}}{q_{x}^2+q_{y}^2+\xi_{0}^{-2}}\\
A(\mathbf{q}_{p},\xi_{p}) &= \frac{W_{p}\xi_{p}^{-1}}{(q_{x}-\pi)^2+(q_{y}-\pi)^2+\xi_{p}^{-2}}.
\end{split}
\end{equation}
By fitting along the diagonal $\mathbf{q}$ direction we can estimate two correlation length scales ($\xi_0$ and $\xi_p$)  and two weights ($W_0$ and $W_p$) for $\mathbf{q}=(0,0)$ and $\mathbf{q}=(\pi,\pi)$ respectively.  An example fit is shown in Fig.~\ref{fig:xi} at $\beta t=4$.  We see that such a model of the data is an extremely good fit and this provides a picture of a very broad pairing mode at $\mathbf{q}=(\pi,\pi)$ and a sharper mode at $\mathbf{q}=(0,0)$.

We analyze the evolution of spectral weight and correlation lengths in Fig.~\ref{fig:xi}(b-g). 
As a function of temperature, shown in Fig.~\ref{fig:xi}(b-c) we find that $\xi_p$ is rather flat and less than one lattice constant while the correlation length for the $\mathbf{q}=(0,0)$ mode is greater than one for $\beta t>3$ and is growing exponentially.  The weight of each mode and their ratio are shown in the inset of Fig.~\ref{fig:xi}(c).  Both $W_0$ and $W_p$ grow as the temperature is decreased while the ratio $W_p/W_0$ is roughly linear until $\beta t=10$.  For temperatures below this, the $W_0$ mode begins to dominate for this interaction strength of $U/t=3$.  This temperature scale coincides with a metal-insulator crossover known to occur here.\cite{fedor:2020}
 We see a continual growth of $\xi_0$ for reduced temperatures, shown in units of the lattice spacing we see that this pairing process remains short range on the scale of $1\to3$ lattice spacings.  This is not the case for $\xi_p$ that remains pinned at values less than one lattice spacing, indicating that this peak is a predominantly local pairing process. 
By fixing temperature $\beta t=4$ we plot similar quantities as a function of chemical potential and again we see $\xi_p$ is independent of $\mu$ while $\xi_0$ increases with doping while its amplitude decays.  While a weak effect at this temperature, this leads to a range of $\mu$ where $\mathbf{q}=(0,0)$ mode dominates $W_p/W_0<1$ and a region where $W_p/W_0>1$ and the $\mathbf{q}=(\pi,\pi)$ mode is dominant.  Nevertheless, at half-filling, we see that the two modes have comparable weight. The doped $W_p/W_0>1$ region also roughly coincides  where the competition  with $p-$ and $d_{xy}-$wave pairing is observed (see appendix~\ref{appendix:mu}).
Interestingly, neither $\xi_0$ nor $\xi_p$ is strongly dependent upon the interaction strength, which we explore in Fig.~\ref{fig:xi}(f-g).  We see, however, that the weight of the modes scales dramatically with $U/t$ values, but expect higher-order corrections to suppress growth beyond $U/t = 3$.

\section{Discussion }\label{main:Disc}  By including a repulsive local Hubbard interaction we have found the existence of positive pairing due to vertex interactions in the weak coupling limit.  We find two dominant momenta for which pairing occurs, the `standard' zero momentum process that we find remains correlated over a short length scale, $\xi_0/a>1$, and also an unexpected $\mathbf{q}=(\pi,\pi)$ pairing of a purely local nature, $\xi_0/a<1$. These observations are robust at intermediate temperatures and exist already in the weak coupling regime at $U/t=2$ that has been heavily studied recently.\cite{Schaefer:2020}  We find that the integrated weight of these two components favours $\mathbf{q}=(0,0)$ at low temperature and half-filling but that contributions from the $(\pi,\pi)$ mode have a substantial contribution to local pairing and dominate at high temperature and away from half-filling. Further, we show explicitly that the dominant positive $d$-wave pairing comes from non-ladder diagrams, a single diagram at second order (with crossed interaction lines) as well as the sum of all 12 third order diagrams. Of important note is that the physical processes described by these diagrams are not the same as those responsible for spin and charge excitations in the particle-hole channel.  We expect that for this model the higher order diagrams will suppress pairing for larger $U/t$ values and low temperatures which will prevent an infinite range superconducting state except at $T=0$ as required by the Mermin-Wagner theorem.  Nevertheless, finite range pairing exists at accessible temperatures and we have shown estimates for the pairing length scales of each mode.

The physical picture that emerges is one of an interaction driven short ranged pairing between two electrons travelling with opposite momenta and spin $k\uparrow$ and $-k\downarrow$ (the $\mathbf{q}=(0,0)$ case) but also rather large contributions from a local collective mode traveling in the $(\pi,\pi)$ direction.  One can visualize this process as the simultaneous hopping of pairs along the diagonal.

We can also speculate as to how our pairing processes relate to insulating and pseudogapped behavior observed in single-particle properties.\cite{fedor:2020,Schaefer:2020}  The predominant theory is that bosonic $\mathbf{q}=(\pi,\pi)$ spin excitations couple with electron quasiparticles to form both the pseudogap and insulating behaviors that are observed in the 2D Hubbard model.  From our results, we conclude that the insulating behavior may also coincide with non-local but finite range $\mathbf{q}=(0,0)$ pair correlations. We would propose that the formation of pairs is not a competitive process with spin-excitations but rather than $\mathbf{q}=(\pi,\pi)$ spin excitations act as the mediator of both $\mathbf{q}=(0,0)$ and $\mathbf{q}=(\pi,\pi)$ pair correlations that might be responsible for opening a gap or pseudogap in the single-particle spectra as long suggested by mean-field RVB singlet models.\cite{yrzreview,leblanc:yrz:raman}
\section{Acknowledgement}
We acknowledge the support of the Natural Sciences and Engineering Research Council of Canada (NSERC) RGPIN-2022-03882 and support from the Simons Collaboration on the Many Electron Problem.

\appendix
\section{Symmetry Channel}\label{appendix:symmetry} 
 The order parameter can be projected into five different symmetry factors based on the irreducible representations of the  $D_{4h}$ symmetry group of the square lattice. For pairing susceptibility, this pertains to applying two symmetry factors belonging to the \emph{same} irreducible representation $\gamma_{\mathbf{k}+\mathbf{q}}$ and $\gamma_{\mathbf{k^\prime}+\mathbf{q}}$ to the incoming and outgoing momenta of a single particle fermionic line as outlined in the main text. The list of symmetry factors used is enumerated in Table~\ref{tab:symmetry}. In the normal state, mixing of the order parameter is forbidden as a component from one irreducible representation cannot mix with another representation.\cite{Annett:1990,Hutchinson_2021} 

 \begin{table}
    \centering
    \begin{tabular}{c|c}
        Symmetry Factor& $\gamma(k)$ \\ \hline \hline
          s & $1$\\ 
          $p_{x}, p_{y}$ & $sin(k_{x}), \,sin(k_{y})$\\
          $d_{xy}$ & $sin(k_{x})sin(k_{y})$\\
          $d_{x^2-y^2}$ & $cos(k_{x}) - cos(k_{y})$\\
    \end{tabular}
    \caption{\label{tab:symmetry} First harmonics of symmetry factor based on the irreducible representation of  $D_{4h}$ group.} 
\end{table}

\section{ Truncated fourth order momentum dependency across symmetry channels} \label{appendix:kcut} 
\begin{figure}
    \vspace{0.15cm}
    \centering
    \includegraphics[width=1\linewidth]{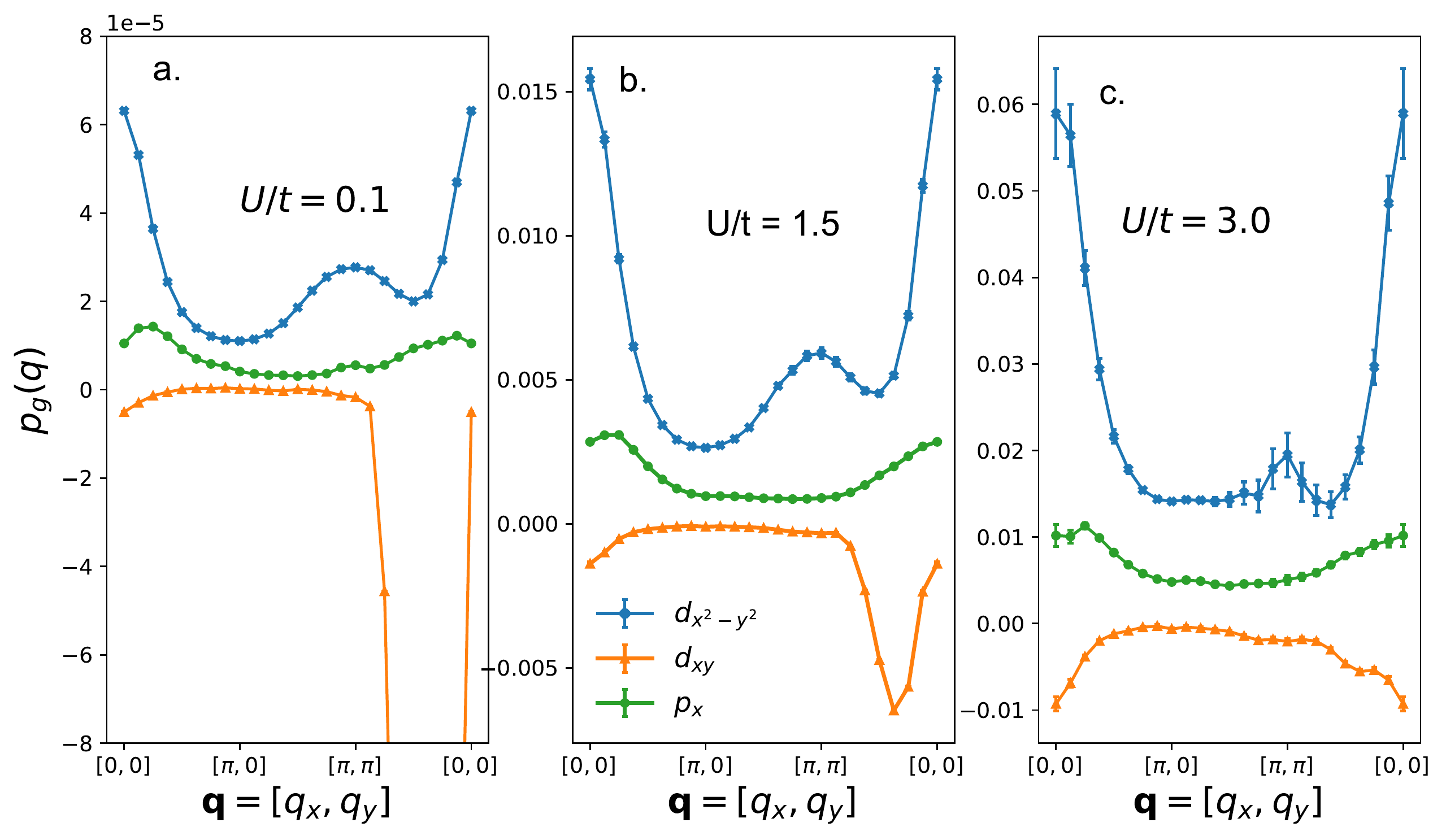}
    \caption{ Truncated fourth order full momentum dependency of $p_{x}$ (green line), $d_{xy}$ (orange line), $d_{x^2-y^2}$ (blue line) pairing at $\beta t = 4$ and a half filling for three different  interaction strength of: (a) $U/t =0.1$, (b) $U/t =1.5$, and (c) $U/t =3$. Sharp negative peak region in panel a. for $d_{xy}$  does not indicate a divergence .}
    \label{fig:Q_comp}
\end{figure}

\begin{figure}
    \centering
    \includegraphics[width=1\linewidth]{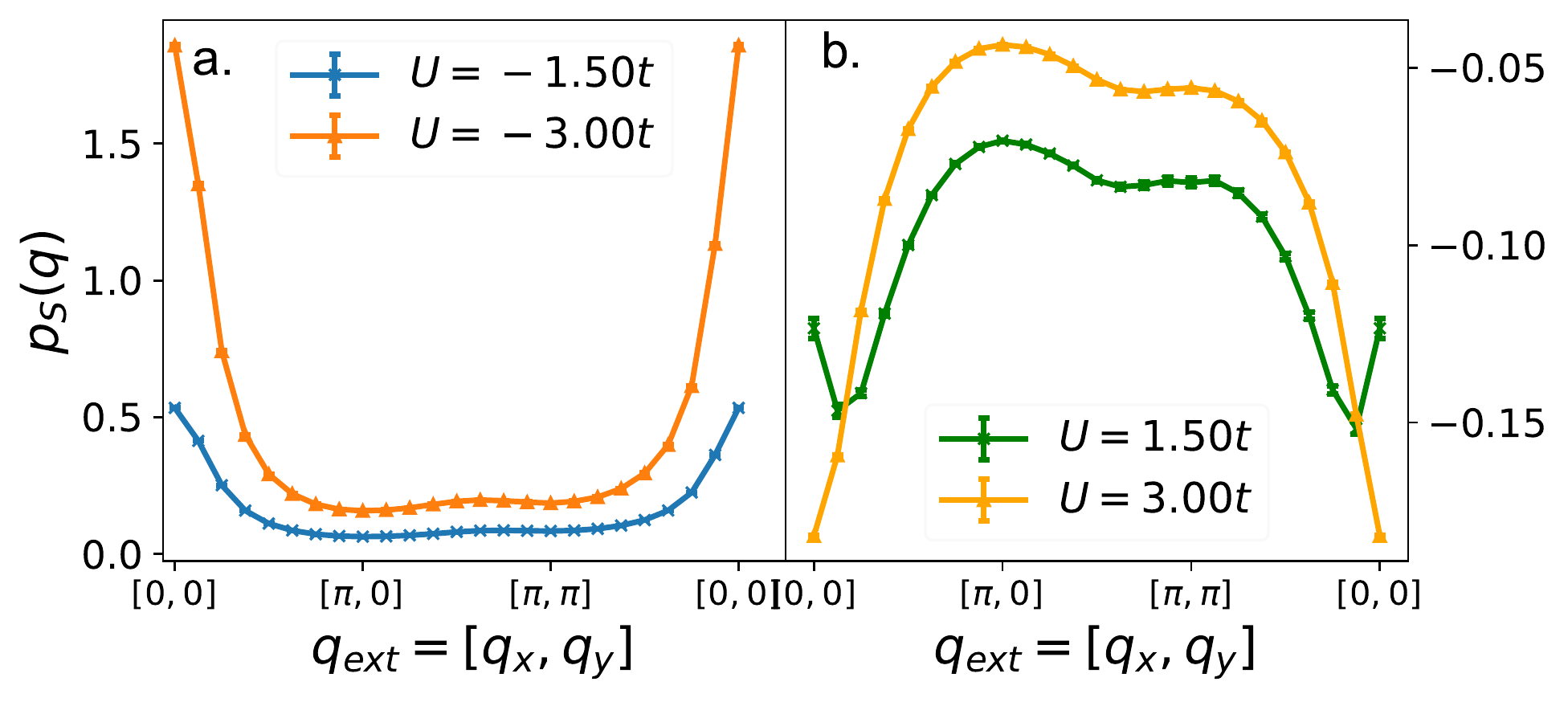}
    \caption{ Truncated fourth order $P_{s}$ along the high symmetry momentum cuts at half-filling  and $\beta t =4$ is presented for (a) attractive  $U=-1.50$ and $U=-3.00t$ interaction. (b) repulsive  $U=1.50$ and $U=3.00t$ interaction. }
    \label{fig:Q_compsw}
\end{figure}

We study the momentum dependence of $P_{g}$ at half filling and $\beta t =4$ in Fig.~\ref{fig:Q_comp} for $p_{x}$, $d_{xy}$, and $d_{x^2-y^2}$ symmetries and in Fig.~\ref{fig:Q_compsw} for $s$-wave. Unlike the main text, we also include fourth-order corrections and see that the qualitative features of $d_{x^2-y^2}$ pairing remain unchanged within the range of interaction strength $U/t=0\to3$ we employ. We find $d_{x^2-y^2}$-wave  pairing to be dominant across the symmetry cuts. Furthermore, the fourth-order correction, despite having a negative contribution to both $\mathbf{q}=(0,0)$ and $\mathbf{q}=(\pi,\pi)$ pairing modes, keeps the overall features reasonably intact  at $U/t=3$ in $d_{x^2-y^2}$ pairing. We emphasize that the two-mode feature is robust at even infinitesimal $U/t$ values since it comes from the leading second-order diagram.  The pairing in $d_{xy}$ channel always remains repulsive at half-filling with a sharp peak centered at $\mathbf{q}=(\pi/3,\pi/3)$. This peak  originates from the ladder diagram in each order and does not mark the breakdown of perturbation. The peak only exists within the weak-coupling limit and gets suppressed by $U/t=3.0$, shown in Fig.~\ref{fig:Q_comp}(c). We find that the $s$-wave pairing in Fig.~\ref{fig:Q_compsw}(a) is negative in the entire momenta space for the repulsive $U/t$ regime. However, in the attractive regime ($U/t<0$ ),  $s$-wave pairing is significantly dominant over other channels and mimics a standard BCS-like picture where a large contribution from $s$ wave $\mathbf{q}=(0,0)$  pairing is expected leading to a zero-center of mass pairing as shown in Fig.~\ref{fig:Q_compsw}(b). The $\mathbf{q}=(0,0)$ $s$-wave  pairing is driven by ladder diagrams with non-ladder diagrams acting to suppress it.

\section{Order-by-Order contribution to $\mathbf{d_{x^2-y^2}}$ pairing} \label{appendix:order_d2} 
\begin{figure}
    \centering
    \includegraphics[width=\linewidth]{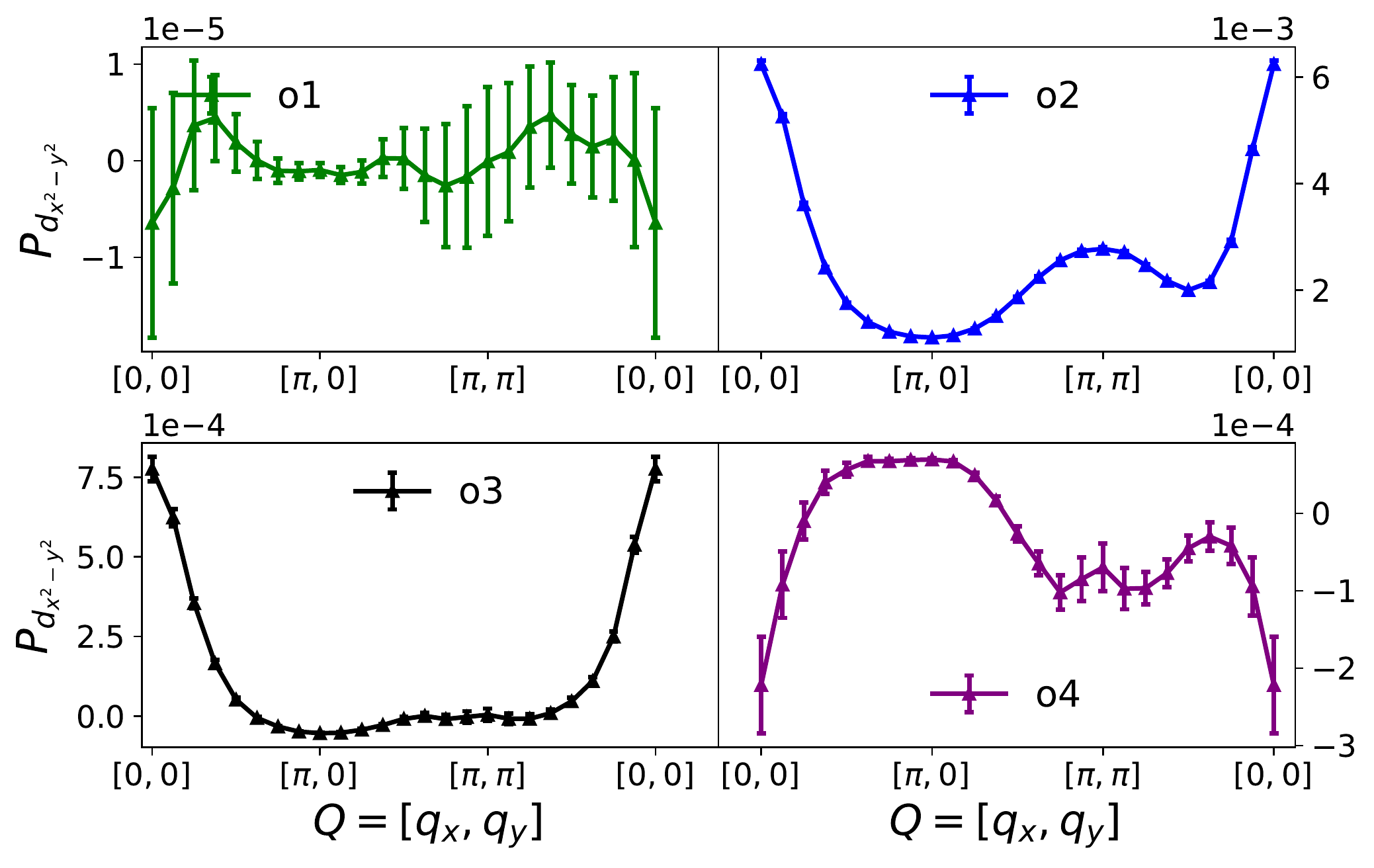}
    \caption{\label{fig:obo_dw}Order by order contribution to $d_{x^2-y^2}$ pairing at $\beta t=4$, $U=3t$ and at half-filling. Order 1 result is zero along the nodal line of $d_{x^2-y^2}-$wave symmetry. }.
\end{figure}

We present order-by-order contributions for high symmetry directions in $q$ for the $d_{x^2-y^2}$ symmetry in Fig.~\ref{fig:obo_dw}.  A standard Hubbard $U$ interaction allows one to split any diagram into a set of particle-particle bubbles or non-ladder diagrams in which the symmetry factor becomes independent of each other. In those ladder diagrams, pairing goes to zero when the scattering momenta vector is along the nodal line of $d_{x^2-y^2}$ symmetry. Consequently, the first-order contribution consisting of a single ladder diagram  is zero along the nodal line, while the region around $\mathbf{q}=(0,\pi)$ has a small but finite negative contribution. 
The second order has two diagrams consisting of a ladder whose contribution is also negligible  and a cross-interacting diagram from which the  $\mathbf{q}=(0,0)$ and $\mathbf{q}=(\pi,\pi)$ modes stem. This one cross-interacting second-order vertex diagram essentially encapsulates the  $d_{x^2-y^2}$ vertex expansion well within the weak coupling limit of $U/t \leq 3$  and truncation order used.  The contribution of this diagram is approximately an order of magnitude higher than all the third-order diagrams combined, the next significant contributor at $U/t=3$.  In fact, the peak at $\mathbf{q}=(\pi,\pi)$ (i.e. equal momentum pairing mode) is a consequence of this crossed interaction diagram with third-order diagrams flat in this region and fourth-ordered diagrams acting to suppress it.

\section{Doping dependency across symmetry channels with second nearest hopping}  \label{appendix:mu} 
\begin{figure}
    \centering
    \includegraphics[width=1\linewidth]{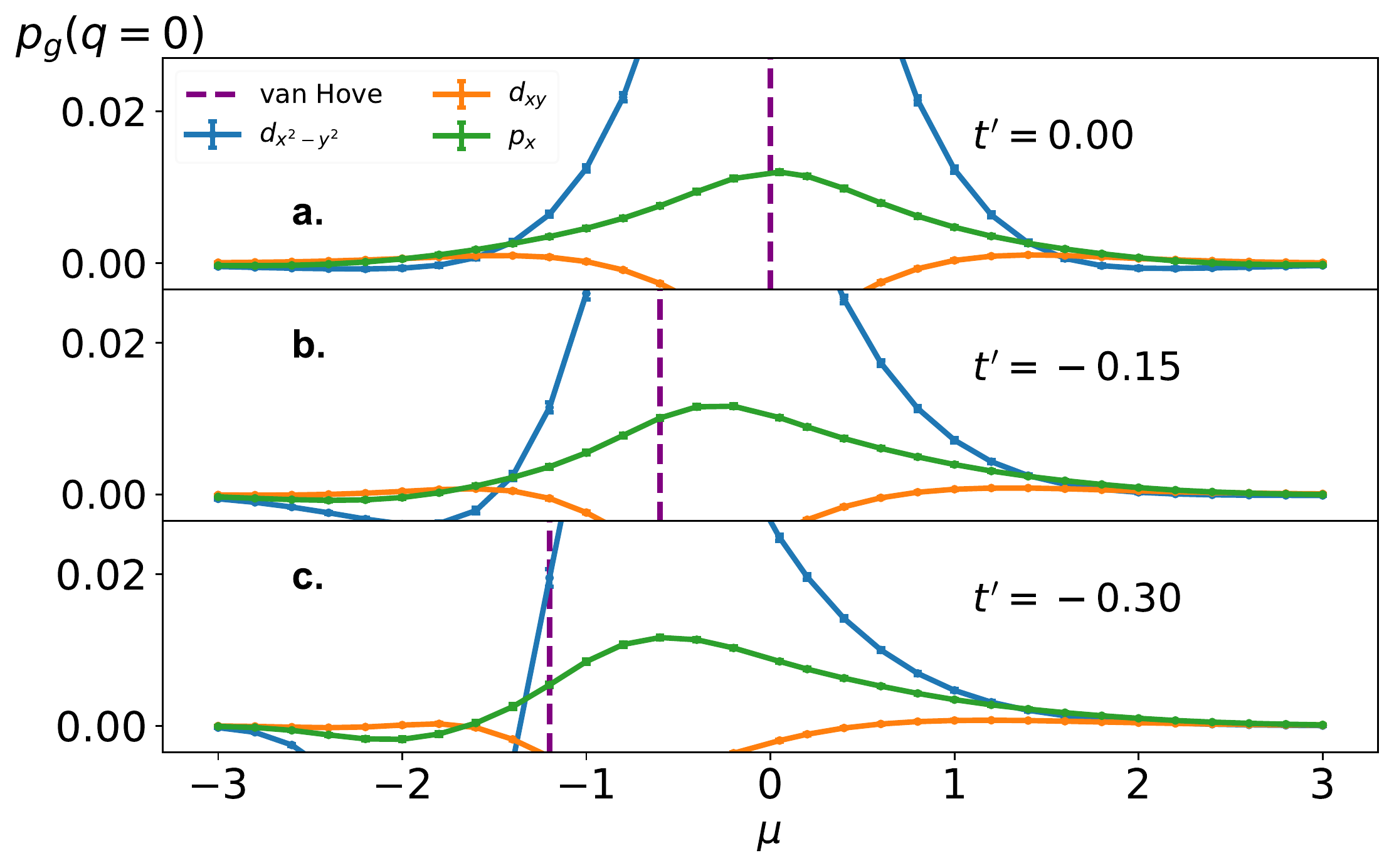}
    \caption{\label{fig:mu_comp} Doping dependence of $\mathbf{q}=[0,0]$ pairing  at $\beta t=4$ and $U=3t$ symmetry factor $p, d_{xy}, d_{x^2-y^2}$ with (a)$t^\prime =0$ (b) $t^\prime= -0.15$ and (c) $t^\prime= -0.30$ }
\end{figure}

The doping dependency of truncated third-order pairing for $p$, $d_{xy}$ and $d_{x^2-y^2}$-wave is studied with the inclusion of the second nearest neighbor hopping $t^\prime$ for $\mathbf{q}=(0,0)$ pairing at $\beta t= 4$ and $U/t =3$ in Fig.~\ref{fig:mu_comp}. For $t^\prime = 0$ case, from the half-filling to intermediate doping strength of $|\mu|<1.4$, the $\mathbf{q}=(0,0)$ $d_{x^2-y^2}$ pairing still remains dominant. However, in the heavily doped regime, $d_{x^2-y^2}$ ceases to be positive and competition between  $p$ and $d_{xy}$ emerges although the pairing strength is comparatively weak. This pairing in this doping regime is highly susceptible to changes in interaction strength, temperature, and higher order correction so no clear phase boundary is established. This competition has already been proven to exist in several perturbative and self-consistent studies. \cite{Romer:2020,Yamada:2002} The inclusion of $t^\prime$  shifts the peaks to hole-doped  and weakens the overall contribution from all symmetry channels. Of an interesting note, Fig.~\ref{fig:mu_comp} shows a remarkable resemblance to 8-site DCA clusters at $U/t=6$ to our infinite system size calculations at $U/t=3$.\cite{chen:2015}


\bibliographystyle{apsrev4-2}
\bibliography{refs.bib}

\end{document}